\documentclass{elsart}

\usepackage{graphicx,amssymb,amsmath,bm}

\journal{Journal of Computational Physics}

\newcommand{\pderiv}[2]{\ensuremath{\frac{\partial #1}{\partial #2}}}
\newcommand{\bra}[2][]{#1\langle #2 #1\rvert}
\newcommand{\ket}[2][]{#1\lvert #2 #1\rangle}

\begin{document}

\begin{frontmatter}

\title{Monte Carlo techniques for real-time quantum dynamics}

\author{Mark R. Dowling}
\ead{dowling@physics.uq.edu.au}

\author{Matthew J. Davis}

\author{Peter D. Drummond}

\author{Joel F. Corney}

\address{ARC Centre of Excellence for Quantum-Atom
  Optics, School of Physical Sciences, University of Queensland,
  Brisbane, Queensland 4072, Australia}


\begin{abstract}
  The stochastic-gauge representation is a method of mapping the
  equation of motion for the quantum mechanical density operator onto
  a set of equivalent stochastic differential equations.  One of the
  stochastic variables is termed the ``weight'', and its magnitude is
  related to the importance of the stochastic trajectory.  We
  investigate the use of Monte Carlo algorithms to improve the
  sampling of the weighted trajectories and thus reduce sampling error
  in a simulation of quantum dynamics.  The method can be applied to
  calculations in real time, as well as imaginary time for which Monte
  Carlo algorithms are more-commonly used.  The method is applicable
  when the weight is guaranteed to be real, and we demonstrate how to
  ensure this is the case.  Examples are given for the anharmonic
  oscillator, where large improvements over stochastic sampling are
  observed.
\end{abstract}


\end{frontmatter}

\section{Introduction}

Complexity is a fundamental problem in theoretical physics
\cite{Feynman1982a}.  It refers to the large size of many physical
systems in terms of their microscopic constituents, and therefore the
near impossibility to predict, from first principles, the detailed
properties of such a system.  In quantum physics the problem is
manifest as the enormous dimension of the appropriate Hilbert space
for most realistic physical systems.  In particular, the calculation
of exact quantum dynamics is notoriously difficult, because the
dimension of the Hilbert space scales exponentially with the number of
modes.  This enormous Hilbert space prevents complete representation
of the evolving many-body quantum state as a numerical state-vector
for large many-body systems. Successful quantum simulation methods can
thus only hope to sample the quantum evolution, to some finite
precision, by use of stochastic methods.

Such quantum Monte Carlo (QMC) techniques have a long history in
first-principles, microscopic calculations of thermal equilibrium and
ground states in quantum
systems~\cite{vonderLinden1992a,Ceperley1999a}.  Certain classes of
QMC methods, such as diffusion or Green's function
approaches~\cite{Ceperley1986a} (projector methods) are restricted to
calculation of ground state properties.  Methods based on path
integrals~\cite{Ceperley1995a} are a simulation through imaginary
time, and can calculate correlations at any temperature.  However,
when it comes to real-time calculations (i.e. dynamics), path integral
QMC methods become difficult because of sign or phase
problems~\cite{Gilks1996a,Landau2000a}.

An alternative approach is provided by phase-space
representations~\cite{Wigner1932a,Drummond1980a,Gardiner1999a}, which
can be used to map quantum dynamics to a set of equivalent stochastic
differential equations.  The number of phase-phase equations scale
polynomially with the number of modes, allowing computationally
tractable simulations.  Phase-space methods have proved useful in the
past for simulating quantum dynamics, particularly in the field of
quantum optics~\cite{Carter1987a,Kinsler1990a,Kinsler1989a} A natural
extension of these techniques is to the field of degenerate quantum
gases, where the interacting particles are atoms or molecules rather
than photons~\cite{Steel1998b,Drummond1999a}.  Recently it has been
discovered that Fermi gases can be treated with related
techniques~\cite{Corney2003a,Corney2004a}.

The mapping of a quantum problem to phase-space equations is far from
unique.  This nonuniqueness can be exploited to tailor the form the
stochastic equation without affecting the physical, ensemble result.
The different choices correspond to different ``stochastic gauges''.
In this paper, we use this freedom to generate stochastic equations
with real weights, which we then sample with Monte Carlo techniques.
The real weights avoid the sign or phase problem encountered in other
QMC approaches to quantum dynamics. Since the stochastic gauge method
is a relatively new technique, we choose to focus here on an
especially simple case with known exact solutions, in order to clarify
the problems and advantages of this real weight approach.

We emphasise, however, that even the simple case of the quantum
anharmonic oscillator that we treat here has highly nontrivial
behaviour when treated as a stochastic problem. It is also relevant to
current experiments on the dynamics of ultra-cold atoms trapped in
optical lattices. These are described rather accurately by the
so-called Bose-Hubbard model, which reduces to the type of single-mode
theory treated here in the Mott-insulator limit in which inter-well
tunneling is suppressed.  We simulate the decay of coherence due to
phase-diffusion, a physical effect that has already been
experimentally observed in recent BEC experiments~\cite{Greiner2002b}.
The present paper focuses on this relatively straightforward and
exactly soluble case, in order to demonstrate the important principles
behind first-principles quantum dynamical simulation in real time. Due
to the linear scaling of these methods with increasing numbers of
modes, we expect that the same basic ideas will apply to multi-mode or
multi-well situations where there are no known exact solutions in
general.

This paper is organised as follows.  In Sec.~\ref{sec: stochgauge} we
review the stochastic-gauge representation~\cite{Deuar2002a} and
motivate the use of Monte Carlo techniques for sampling the weighted
stochastic trajectories.  In Sec.~\ref{sec: anharm} we introduce a
simple model for interacting quantum dynamics, and describe the design
of a gauge that results in real weights.

In Sec.~\ref{sec: metropolis} we review the Metropolis-Hastings
algorithm~\cite{Metropolis1953a,Hastings1970a,Chib1995a} and describe
its application to stochastic-gauge simulations, demonstrating the
improvements over the usual stochastic sampling for the same type of
gauge.

In Sec.~\ref{sec: branching} we describe an alternative scheme, based
on stochastic sampling, but with a branching algorithm for efficient
handling of the weights.

Finally, in Sec.~\ref{sec: conclusion} we discuss the advantages and
disadvantages of each Monte Carlo method in the context of future
applications.

\section{The Stochastic-Gauge Representation}

\label{sec: stochgauge}

The \textit{stochastic-gauge} (or \textit{gauge-$P$}) representation
is a generalisation of the positive-$P$ ($+P$) representation
\cite{Drummond1980a,Gardiner1999a}, where the density operator is
expanded as a positive distribution over an off-diagonal,
over-complete basis set of coherent states.  The essential difference
between the $+P$ and the stochastic-gauge representation is that the
later is defined over a quantum phase space with an additional
dimension termed the \textit{weight} $\Omega$, such that the total
phase space vector is $\vec{\alpha} = (\bm{\alpha}, \bm{\beta},
\Omega)$ of complex dimension $2M+1$.  Throughout this paper the $+P$
variables $(\bm{\alpha},\bm{\beta})$ are referred to as \textit{mode
  variables}.  For a complete description of the method we refer the
reader to Ref.~\cite{Deuar2002a}, however we briefly summarise the
main features of the method below.

The procedure for calculating quantum dynamics using the
stochastic-gauge representation is similar to that using the $+P$
representation and involves deriving, via a Fokker-Planck equation, a
set of stochastic differential equations equivalent to the original
quantum master equation.  The most general quantum-dynamical evolution
may be written as a master equation of the form
\begin{equation}
\pderiv{\hat{\rho}}{t} = L[\hat{\rho}],
\end{equation}
were $L$ is a Liouville superoperator (e.g.\ $L[\hat{\rho}] = -i
[\hat{H},\hat{\rho}]$ for unitary evolution).  To calculate the
quantum dynamics using the stochastic-gauge representation we expand
the density operator in an over-complete basis set as
\begin{equation}
\hat{\rho} = \int d^{4M+2} \vec{\alpha} G(\vec{\alpha},\vec{\alpha}^*) \hat{\Lambda}(\vec{\alpha}),
\end{equation}
where
$$\hat{\Lambda}(\vec{\alpha}) = \Omega || \bm{\alpha}\rangle\langle
\bm{\beta}^*|| \exp[-\bm{\alpha} \cdot \bm{\beta}],$$
is the kernel,
$G(\vec{\alpha},\vec{\alpha}^*)$ is the (non-unique) gauge
distribution function, and
\begin{equation}
||\alpha \rangle = \sum_n \frac{\alpha^n}{\sqrt{n!}} \ket{n},
\end{equation}
is a Bargmann coherent state \cite{Gardiner1999a}.

By use of operator identities for creation and annihilation operators
acting on the kernel and subsequent integration by parts (with the
assumption that boundary terms vanish), it is possible to show that
any master equation involving only two-body terms is equivalent to a
gauge distribution function evolving according to a positive-definite
Fokker-Planck equation.  The dynamical moments may therefore be
obtained by evolving an equivalent set of stochastic differential
equations (SDEs) and taking stochastic averages of an appropriate
product of stochastic variables~\cite{Gardiner1985a}.

At this stage it is possible to add arbitrary terms to the
Fokker-Planck differential operator that give zero when acting on the
kernel.  These terms cannot affect the quantum averages, but add
stochastic gauges --- arbitrary functions on phase space --- to the
drift part of the corresponding SDEs~\cite{Deuar2002a}.  The central
result is easy to state.  For an $M$ mode quantum system with $+P$ Ito
equations of the form
\begin{equation}
\dot{\vec{\alpha}} = \vec{A}^{(+)} + \mathbf{B}^{(+)} \vec{\xi},
\end{equation}
where $\vec{A}^{(+)}$ and $\mathbf{B}^{(+)}$ are the positive-$P$
drift vector and diffusion matrix, respectively, then the
stochastic-gauge equations for the system are
\begin{eqnarray}
\label{eq: itogaugevecalphadot} \dot{\vec{\alpha}} &=& \vec{A}^{(+)} 
 + \mathbf{B}^{(+)} \left( \vec{\xi} -\vec{g}   \right) \, , \\
\label{eq: itogaugeomegadot} \dot{\Omega} &=&  \Omega \vec{g}^T \vec{\xi}\, . 
\end{eqnarray}
Here $\vec{g} = \vec{g}(\vec{\alpha})$ is a vector of
\textit{stochastic gauges} and $\vec{\xi}$ is a vector of gaussian noises
where
\begin{equation}
\langle \xi_i (t) \xi_j (t')\rangle =\delta_{ij} \delta (t-t')\, .
\end{equation}
The stochastic gauges $\vec{g}$ can  be used to modify the deterministic
evolution of the stochastic trajectories and are therefore called
\textit{drift gauges}.  

Also possible are \textit{diffusion gauges}
which arise from the non-unique factorisation of drift matrix $ \mathbf{D}$
appearing in the Fokker-Planck equation into a noise matrix $ \mathbf{B}$ for
the SDEs, where
\begin{equation}
\mathbf{D} = \mathbf{B} \mathbf{B}^T \, \, .
\end{equation}
Diffusion gauges can be used, for example, to
``squeeze'' the noise between stochastic variables to improve sampling
\cite{Plimak2001a}.  More general types of diffusion gauge are
possible in the full stochastic-gauge formalism~\cite{Deuar2002a,Deuar2004a}.

For a single mode, quantum-dynamical averages of normally ordered
products of creation and annihilation operators are calculated as
stochastic averages in the following manner
\begin{equation}
\label{eq: gaugePquantstochav} \langle \left( \hat{a}^\dagger \right)^m \hat{a}^n
\rangle_{\mathrm{QM}} = \frac{\langle \Omega \beta^m \alpha^n + (\Omega
  \beta^n \alpha^m)^*\rangle_{\mathrm{stoch}}}{\langle \Omega + \Omega^*\rangle_{\mathrm{stoch}}}.
\end{equation}
In principle any gauge that does not introduce boundary terms
on partial integration will reproduce the exact quantum averages in the limit
that an infinite number of trajectories are simulated,
and so all gauges represent the same physics.  However, in practice we
may only simulate a finite number of trajectories due to limited
computing resources, and so we would like to choose a gauge that gives
rise to the most compact phase space distribution
$G(\vec{\alpha},\vec{\alpha}^*,t)$ possible.  A narrower distribution
means that fewer stochastic trajectories need to be sampled to obtain
quantum averages with a given accuracy.  The situation is similar to
classical or quantum electrodynamics, where a judicious choice of
gauge simplifies the solution of certain problems.

\subsection{Monte Carlo techniques for the weight variable}

\label{sec: mcweight}

\begin{figure}
\begin{center}
  \centerline{\includegraphics[width=120mm]{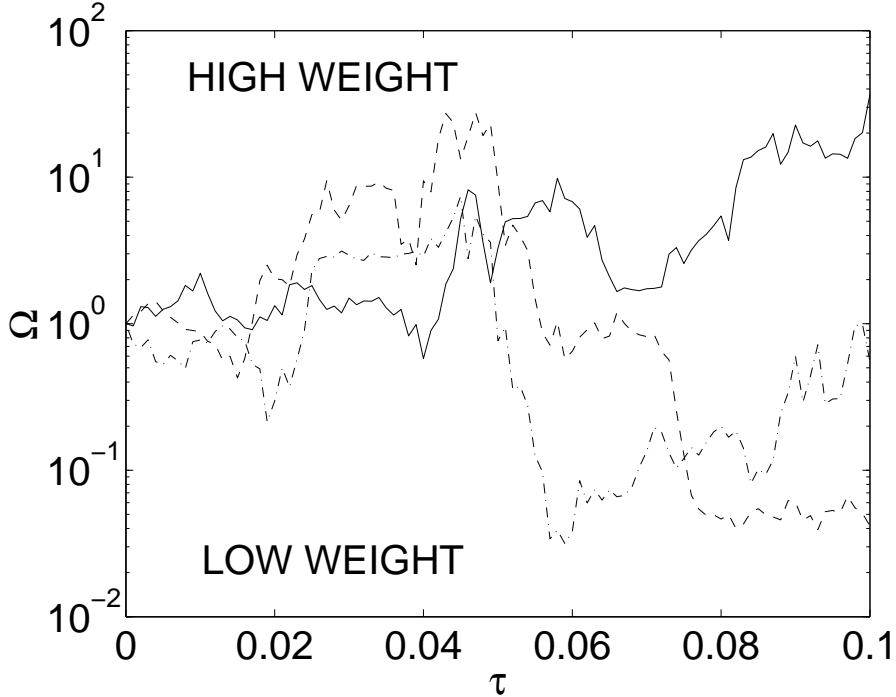}}
\caption{
  Illustration of the spreading of weights in a stochastic gauge
  simulation for the anharmonic oscillator.  These three trajectories
  of $\Omega$ were generated using randomly chosen noises for a
  real-gauge simulation of the anharmonic oscillator ($A = 2$,
  $\lambda = 1/2$), as described in Sec.~\ref{sec: gaugechoice}, with
  mean atom number $\bar{n} = 100$.  The total simulation time was
  $\tau_{\rm tot} = 1/\sqrt{\bar{n}} = 0.1$ and the time step used was
  $\Delta \tau = \tau_{\rm tot} / 10^3 = 10^{-4}$.  Note that the
  relative weights of the trajectories vary in time so the Metropolis
  algorithm can only be targeted to sample the distribution at a
  single chosen time.  The branching algorithm clones and kills
  trajectories continuously in time in proportion to their weight so
  as to obtain continuous-time samples of the distribution.}
 \label{fig: weightspread}
\end{center}
\end{figure}

The motivation for this work is that in the stochastic gauge
representation, stochastic trajectories are often generated with weights
that vary over many orders of magnitude as illustrated in
Fig.~\ref{fig: weightspread}. From Eq.~(\ref{eq: gaugePquantstochav})
we can see that the trajectories with relatively high weight
contribute more to the overall stochastic averages than those with
relatively low weight.  
 
The situation is analogous to path-integral calculations of quantum
averages in imaginary time (thermal quantum averages), where quantum
Monte Carlo techniques have long been used to sample the weight
parameter more efficiently~\cite{Pollock1984a,Ceperley1986a}.  In the
standard stochastic gauge prescription the stochastic trajectories are
generated randomly.  If the weight parameter can be interpreted as a
probability, then this suggests that more sophisticated Monte Carlo
techniques can be used to efficiently sample high-weight trajectories,
thus improving the sampling of physical averages.  In this paper we
focus on real-time dynamics, although the same techniques apply to
imaginary-time stochastic gauge calculations~\cite{Corney2004a}.
Previously Monte Carlo techniques have had limited success with
real-time quantum dynamics, where path-integral approaches are plagued
by sign problems due to the rapidly-oscillating phase
\cite{Ceperley1999a,Egger1994a}.

However, there is a complication --- in the stochastic-gauge
representation the weight is complex in general.  In principle it is
possible to apply Monte Carlo techniques to problems with
a complex weight --- by treating the modulus of the weight as the
`importance'. In practice, however, sign problems are encountered when the
phase of the weight eventually becomes evenly distributed around the
unit circle in the complex plane.  In order to strictly interpret the
weight as a probability distribution, we must ensure that our choice
of gauge leads to a weight parameter that is real.  This means that
the gauge functions we introduce must also be real.

\section{Example: The Kerr Anharmonic Oscillator}

\label{sec: anharm}

In this section we introduce a single-mode boson model for quantum
nonlinear dynamics --- the Kerr anharmonic oscillator.  The
Hamiltonian is
\begin{eqnarray}
\label{eq: anharmH} \hat{H} = \hbar \omega_0 (\hat{a}^\dagger \hat{a} + \frac{1}{2}) + \frac{1}{2}\kappa \hat{a}^{\dagger 2} \hat{a} ^2 = \hbar \omega_0 (\hat{n}+\frac{1}{2}) + \frac{1}{2}\kappa \hat{n} (\hat{n} - 1),
\end{eqnarray}
where $\hat{n} = \hat{a}^\dagger \hat{a}$ is the number operator.
Traditionally this Hamiltonian has been used to describe the Kerr
effect in nonlinear optics. It has received
renewed interest recently, as it is the restriction to a single site
of the Bose-Hubbard model.  This has been shown to describe ultra-cold
bosonic atoms in an optical lattice~\cite{Jaksch1998b}, a topic of
recent theoretical and experimental interest~\cite{Greiner2002a,Greiner2001a}.

Throughout this paper we work in the interaction picture with
\begin{equation}
H_{\mathrm{int}} = \frac{1}{2}\kappa \hat{a}^{\dagger 2} \hat{a} ^2.
\end{equation}
A number state is an eigenstate of this Hamiltonian.
Introducing
the dimensionless time $\tau = \kappa t/\hbar$, a number state
 evolves according to
\begin{equation}
\ket{n(\tau)} = \exp(-i  n (n-1) \tau/2 )\ket{n(0)}
\end{equation}
Hence, any initial state can be decomposed into number states and an
exact solution found.  In particular, for an initial coherent state
$\ket{\psi(0)} = \ket{\alpha}$ with $\alpha$ real, the solution is
\begin{equation}
\ket{\psi(\tau)} = e^{- |\alpha|^2/2} \sum_n
\frac{\alpha^n}{\sqrt{n!}} \exp(-i n(n-1) \tau/ 2) \ket{n},
\end{equation}
As this model both has an exact solution and includes the nonlinearity
that is a feature of models such as the Bose-Hubbard Hamiltonian, it
forms an excellent testing ground for quantum simulation
methods~\cite{Dowling2005a}.

An important quantum feature of this Hamiltonian is that given an
initial coherent state $\hat{\rho}(0)=\ket{\alpha}\bra{\alpha}$ with
mean boson number $\bar{n} = |\alpha|^2$, the dynamics display a
series of collapses and revivals.  Defining the quadrature variables
\begin{equation}
\hat{X} = (\hat{a}+\hat{a}^\dagger)/2, \quad \hat{Y} = (\hat{a}-\hat{a}^\dagger)/2i,
\end{equation}
we find there are three characteristic timescales for anharmonic
oscillator dynamics.  The quadratures initially undergo oscillations
with period $\tau_{\mathrm{osc}} \sim O(1/\bar{n})$, which are damped
to zero over a time of $\tau_{\mathrm{coll}} \sim
O(1/\sqrt{\bar{n}})$.  However, the oscillations revive at time
$\tau_{\mathrm{rev}} \sim O(1)$, which for a large mean boson number
can be many times the collapse time.

Because the period of oscillation of the quadratures can be very short
for large mean atom number, and because we are most interested in the
envelope of the collapse, we choose to perform the calculations in a
rotating frame.  The angular frequency of the rotation is equal to the
mean atom number, and the $X$-quadrature, whose exact solution is
\begin{equation}
 \langle \hat{X}(\tau) \rangle = \sqrt{\bar{n}} e^{\bar{n} (\cos(\tau) - 1)} \cos( \bar{n} (\sin( \tau)- \tau)) \, \, ,
\end{equation}
collapses dynamically for large mean atom number according to:
\begin{equation}
 \langle \hat{X}(\tau) \rangle \simeq \sqrt{\bar{n}} e^{- \bar{n}  \tau^2} \,  .
\end{equation}
For the anharmonic oscillator Hamiltonian the stochastic-gauge
Stratonovich SDEs are
\begin{eqnarray}
\dot{\alpha} &=& - i \alpha^2 \beta + i  \alpha / 2 + \sqrt{- i } \alpha \cosh(A) (\xi_\alpha - g_\alpha) \\
\dot{\beta} &=&  i  \beta^2 \alpha + i  \beta / 2 + \sqrt{ i } \beta \sinh(A) (\xi_\beta - g_\beta) \\
\dot{\Omega} &=&  \Omega \sum_{j=\alpha,\beta} g_j \xi_j \, .
\end{eqnarray}
Here $A$ is a diffusion gauge that we choose to be constant in this
paper, $g_j$, $j=\alpha,\beta$ are drift gauges to be chosen
subsequently, and $\xi_j$ are dimensionless Gaussian noise terms
chosen so that:
\begin{equation}
\langle \xi_i (\tau) \xi_j (\tau')\rangle =\delta_{ij} \delta (\tau-\tau')\, .
\end{equation}

For better numerical performance we choose to work with \textit{log
  variables}:
 \begin{eqnarray}
 \theta = (1/2) \log(\alpha\beta), \quad \phi = (1/2i) \log(\alpha/\beta), \quad \omega = \log(\Omega).
 \end{eqnarray}
  These obey the following Stratonovich SDEs:
\begin{eqnarray}
\label{eq: anharmthetadot} \dot{\theta} &=& \frac{1}{2} e^{-A}(\xi_1-g_1 - i (\xi_2-g_2) ), \\
\label{eq: anharmphidot} \dot{\phi} &=& - e^{2 \theta} + \frac{1}{2} - \frac{1}{2} e^{A} (\xi_1-g_1 + i (\xi_2-g_2)), \\
\label{eq: anharmomegadot} \dot{\omega} &=& S_\omega + \sum_{j=1,2} g_j \xi_j,
\end{eqnarray}
where we have defined the linearly transformed noises $\xi_1 =
(\xi_\alpha + \xi_\beta)/\sqrt{2}$ and $\xi_2 = (\xi_\alpha -
\xi_\beta)/\sqrt{2}$, which obey the same statistics, and similarly
$g_1 = (g_\alpha + g_\beta)/\sqrt{2}$ and $g_2 = (g_\alpha -
g_\beta)/\sqrt{2}$.  The term $ S_\omega$ is a Stratonovich correction
factor which depends on the gauge choice, and will be calculated for
each specific case. We note here that, unlike in the corresponding
classical oscillator equations, both $\theta=\theta_X+i\theta_Y$ and
$\phi=\phi_X+i\phi_Y $ are intrinsically complex.

\subsection{Choice of Gauge}

\label{sec: gaugechoice}

In this section we discuss possible choices of diffusion gauge for the
anharmonic oscillator with a view to using Monte Carlo techniques to
sample the weight.

In Ref.~\cite{Drummond2003a} Drummond and Deuar investigated the following
drift-gauge choice for the anharmonic oscillator
\begin{equation}
g_1 = i g_2 =  i e^{A+2 \theta_X} \sin(2 \theta_Y),
\label{eqn:complexg}
\end{equation}
where the $X$ and $Y$ superscripts refer to the real and imaginary
parts of the phase-space variable respectively (this notation is used
throughout this paper).  They found that this gauge extended
simulation times by many orders of magnitude for the same number of
stochastic trajectories.  In particular they could simulate well past
the collapse time for an initial coherent state.

As noted in Sec.~\ref{sec: mcweight}, for the Monte Carlo methods to
be successful it is highly desirable to have a real weight.
Unfortunately the above choice of gauge is complex and thus leads to
complex weights.  We have trialled the use of Monte Carlo techniques
based on using the modulus of the weight, but found these to be
unsuccessful due to the phase problem described in the introduction.

In order to have real weights, we would like to design a real gauge
with similar properties to the above complex gauge to give a similar
extension of simulation time.  The reason that the complex gauge
Eq.~(\ref{eqn:complexg}) improves simulation times is that it removes
a driving term from the imaginary part of the $+P$ $\phi$ equation.
The presence of this term forces trajectories to diverge to infinity
in a non-classical direction in phase-space ($\phi_Y \rightarrow \pm
\infty$), thus resulting in large sampling errors.  At the same time
the equations for $\theta$ and the real part of $\phi$ are unaffected
due to a cancellation between $g_1$ and $g_2$.  We could remove the
offending term from the $\phi_Y$ equation with a single real gauge
$g_2$, however this would necessarily appear in the $\theta_Y$
equation leading to an instability that causes poor sampling.

Thus it seems that a drawback to choosing real gauges is that there is
substantially less control over the stochastic trajectories.  This
could have been anticipated because, in general, with a complex gauge we
have as many gauge degrees of freedom as we do real phase space
dimensions, $4M$ (excluding the weight dimensions).  However with real
gauges we only have half as many gauge degrees of freedom, $2M$,
meaning that we cannot independently control the drift in the real and
imaginary components of each of the mode variables.

An additional source of gauge freedom is obtained from choosing a
non-square noise matrix in deriving the SDEs.  This freedom was
pointed out in~\cite{Deuar2002a} but not explored.  Specifically, it
is possible to take the noise matrix to be of the general form
\begin{equation}
\mathbf{B}= [\mathbf{B}_0, \mathbf{Q}],
\end{equation}
where $\mathbf{B}_0$ is a square ($2M \times 2M$) noise matrix such
that $\mathbf{B}_0\mathbf{B}_0^T = \mathbf{D }- \mathbf{Q}
\mathbf{Q}^T$, and $\mathbf{Q}$ is a $2M \times W$ matrix whose
entries are arbitrary complex functions.  This choice reproduces the
correct moments in the limit of a large number of stochastic
trajectories, but introduces more than the minimum number of noise
terms into the stochastic equations.  Naively this could be expected
to lead to worse sampling errors.  However, this additional gauge
freedom allows us to overcome the restrictios of the standard real
gauges and improve the sampling overall.

For the anharmonic oscillator we choose
\begin{equation}
\mathbf{Q} = \begin{bmatrix} 0 & 0 \\ \lambda & i \lambda \end{bmatrix},
\end{equation} 
in the $(\theta, \phi)$ variables, where $\lambda$ could be an
arbitrary complex function on phase space, although we choose it to be
constant in our example.  Here $\mathbf{Q}$ satisfies
$\mathbf{Q}\mathbf{Q}^T = 0$ so the other noise terms are unaffected.
In doing so we have added a term of the form $\lambda(\xi_3 + i
\xi_4)$ to the equation of motion for $\phi$, Eq.~(\ref{eq:
  anharmphidot}).  An equivalent way of understanding this additional
noise is that due to the analytic nature of the stochastic gauge
kernel we have
\begin{equation}
\lambda \left(\frac{\partial^2}{\partial \phi_X^{ 2}} + 
\frac{\partial^2}{\partial \phi_Y^{2}} \right) \Lambda(\vec{\alpha}) = 0.
\end{equation}
Thus we are free to add $\lambda (\partial^2 / \partial \phi_X^2 +
\partial^2 / \partial \phi_Y^2)$ to the Fokker-Planck differential
operator without affecting the physical moments.  

We are now able to introduce additional gauges to the $\dot{\phi}$
equation in the manner described in~\cite{Deuar2002a} so that the
extra term in the $\dot{\phi}$ equation becomes
\begin{equation}
\lambda((\xi_3-g_3) + i (\xi_4-g_4)),
\end{equation} 
and the gauges $g_3$ and $g_4$ enter the weight equation in the same
way as the other drift gauges.  The extra noise allows us to control
the $\phi_Y$ divergence using only a real gauge, without affecting the
$\theta_Y$ equation.  Specifically we choose
\begin{eqnarray}
g_1 &=& g_2 = g_3 = 0, \\
\label{eq: realgaugechoice} g_4 &=& - \frac{e^{2 \theta_X}}{\lambda} \sin(2 \theta_Y). 
\end{eqnarray}

The final SDEs that we focus on sampling for the rest of this paper
are summarised as
\begin{eqnarray}
\label{eq: realgaugethetadot} \dot{\theta} &=& \frac{1}{2} e^{-A}(\xi_1 - i \xi_2 ), \\
\nonumber \dot{\phi} &=& - e^{2 \theta_X} \cos(2 \theta_Y) + \frac{1}{2} \\
\label{eq: realgaugephidot} &&- \frac{1}{2} \left( e^{A} (\xi_1 + i \xi_2) - 2 \lambda(\xi_3 + i \xi_4) \right), \\
\label{eq: realgaugeomegadot} \dot{\omega} &=& S_\omega - \frac{e^{2 \theta_X} }{\lambda} \sin(2 \theta_Y) \xi_4,
\end{eqnarray}
where $S_\omega = - e^{4 \theta_X} \sin^2(2 \theta_Y) / 2 \lambda^2$ is
the Stratonovich correction in the weight equation.  Note that
although we choose $g_3 = 0$ it is still necessary to include the
noise $\xi_3$ for the mapping to be exact.

Of course there are other possible choices of real gauges, but we have
found this combination to be well-suited to illustrating the
improvements possible with Monte Carlo sampling.

\begin{figure}[]
  \centerline{\includegraphics[width=80mm]{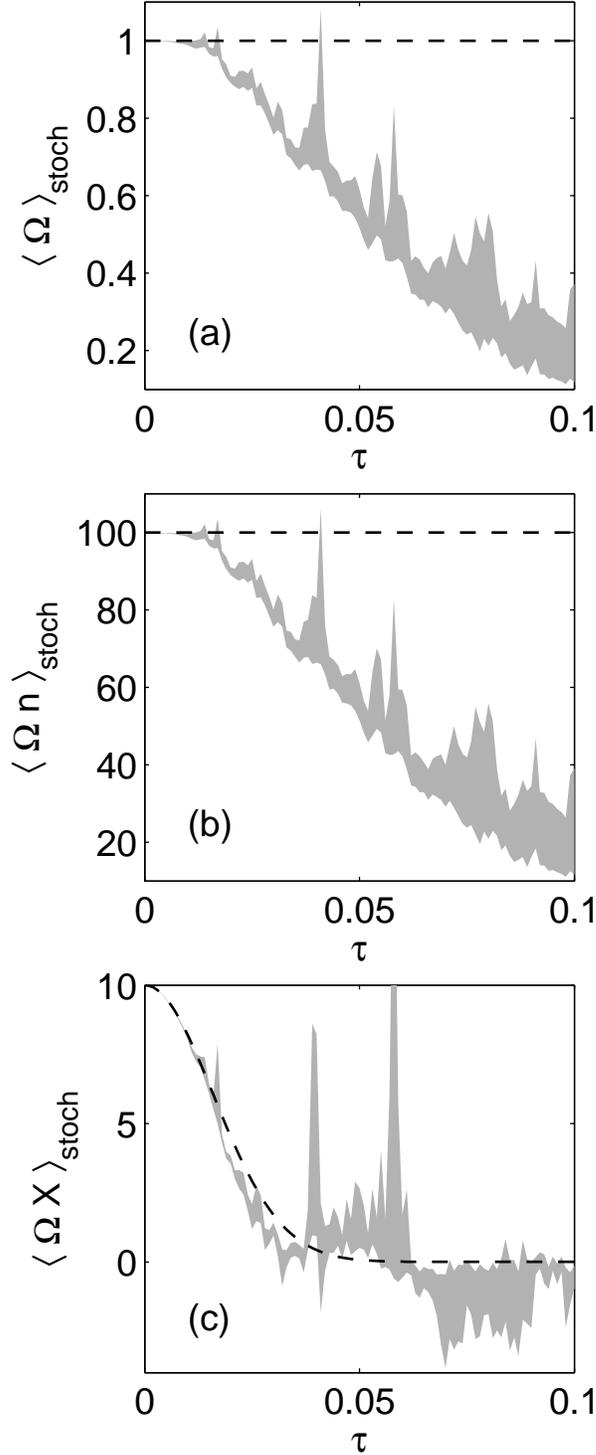}}
  \caption{
  Real-gauge simulation of anharmonic oscillator dynamics for an
  initial coherent state with mean atom number $\bar{n} = 100$ ($A =
  2$, $\lambda = 1/2$).  (a) Mean weight, (b) mean atom number
  (corresponding stochastic variable $n = \alpha \beta$) and (c) mean
  $X-$quadrature (corresponding stochastic variable $X = (\alpha +
  \beta)/2$).  Averages were carried out over $10^6$ stochastic
  trajectories.  The total simulation time was $\tau_{\rm tot} = 1 /
  \sqrt{\bar{n}} = 0.1$ and the time step used was $\Delta \tau =
  \tau_{\rm tot} / 10^3 = 10^{-4}$.  The shaded region indicates the
  estimated error in the simulation and the stochastic means are in
  the centre of the region.  The analytic results are shown as dashed
  lines.  The sampling is poor except for short times due to the
  skewed weight distribution.}
\label{fig: realgaugestoch}
\end{figure}

In Fig.~\ref{fig: realgaugestoch} we illustrate stochastic sampling of
the above SDEs.  We begin with an initial coherent state with mean
atom number $\bar{n} = 100$, and simulate to a final time of $\tau = 1
/\sqrt{\bar{n}} = 0.1$, which is of the order of the collapse time.
Clearly the sampling of the solution is only accurate for short times.  In
particular, the sampled $\langle \Omega \rangle_{\mathrm{stoch}}$
decays towards zero when it should remain at one for all times in a
simulation of unitary dynamics.  The error in sampling the mean atom
number, which should remain at 100, is due almostly entirely to poor
sampling of the weights, as can be seen by the almost identical decay.
Even worse, the estimated error in the means is small despite the fact
that they are clearly far from the analytic result.  This poor
sampling occurs more quickly on the scale of the collapse for larger
mean atom number.

This behaviour can be understood by considering the nature of the
distribution of weights.  We find that the weight parameter evolves
with time towards zero for most trajectories.  However, we know that
the mean weight must be one, and so there must exist a small number of
trajectories with large weights.  Thus the distribution of weights
seems to be skewed towards zero with a long tail extending to large
weights.  Such a distribution is difficult to sample, so it is not
surprising that we underestimate the errors in the means as these
assume Gaussian statistics.  Fortunately, Monte Carlo techniques are
particularly successful in sampling skewed distributions. Below we
describe the application of the Metropolis-Hastings algorithm and a
Monte Carlo branching algorithm to our SDEs for the anharmonic
oscillator.

\section{The Metropolis-Hastings Algorithm}

\label{sec: metropolis}

The Metropolis-Hastings algorithm is a well-known technique for
generating samples of a multi-dimensional probability distribution.
Here we briefly describe the algorithm and discuss how to estimate
errors in quantities derived from such samples.  In particular this
may be useful for readers from the ultra-cold matter community who may
be unfamiliar with these methods.

\subsection{Algorithm}

The Metropolis algorithm was first described in
1953~\cite{Metropolis1953a} and was used to sample a thermal Boltzmann
distribution, which can be viewed as a
probability distribution over the space of all thermally accessible
states of a system.  Subsequently the Metropolis algorithm has been
generalised to many types of probability distributions/densities over
both discrete and continuous spaces, and there is a vast amount of
mathematical literature on Markov Chain Monte Carlo (MCMC)
techniques (see e.g. \cite{Gilks1996a,Landau2000a}.)

In this section we summarise the Metropolis algorithm in a more
general form due to Hastings~\cite{Hastings1970a}.  A well-written
introduction to the Metropolis-Hastings algorithm may be found in
\cite{Chib1995a}, and here we follow their notation.  Formally, the
Metropolis algorithm is a MCMC technique for efficiently sampling a
probability distribution, $\pi(s)$, where $s \in S$ represents the
state of some system, and $S$ is the domain of the distribution known
as the state space.  To be a true probability distribution
$\pi(\cdot)$ must be normalised $\int ds \pi(s) = 1$, however one of
the virtues of the algorithm is that the functional form of the
distribution need only be known up to a constant factor in order to
apply the algorithm\footnote{e.g.\ the partition function $\mathcal{Z}
  = \sum_{s \in S} e^{-E(s)/ k_B T}$, where $E(s)$ is the energy of
  the system in state $s$, need not be know in order to sample the
  Boltzmann distribution, where the probability that the system is in
  state $s$ is given by $P(s) = e^{-E(s)/k_B T}/\mathcal{Z}$}.

A Markov chain can be thought of as a random walk through state space
where the probability of making a particular step depends only on the
current location.  It is a sequence of points, $s_1, s_2, \ldots s_n
\in S$, where the $s_i$ are random variables such that
\begin{equation}
p(s_i|s_1,s_2 \ldots s_{i-1}) = p(s_i|s_{i-1}).
\end{equation}
The conditional density $p(s_i|s_{i-1})$ is known as the
\textit{transition density} of the Markov chain.  If there exists an
\textit{invariant density} 
\begin{equation}
\pi^*(s) = \int ds' \pi^*(s') p(s|s'),
\end{equation}
that satisfies the condition of \textit{detailed balance}
\begin{equation}
p(s'|s) \pi^*(s)  =  p(s|s') \pi^*(s'),
\end{equation}
then it may be shown that the Markov chain \textit{converges} to the
distribution $\pi(\cdot)$.  More precisely, the elements of the chain
$\{s_i | i = b+1,\ldots, n \}$ are unbiased samples from the
distribution $\pi^*(\cdot)$ to within some given precision, where $b
\geq 0$ is known as the \textit{burn-in} and represents the number of
steps required for the chain to converge to within that precision.

The Metropolis-Hastings algorithm solves the problem of determining
what transition density to use so as to generate a given invariant
distribution.  It is concerned with generating samples from some
\textit{target density} $\pi(\cdot)$, known apart from a constant
factor, and does so by determining a suitable transition density for a
Markov chain to converge to this distribution.  To do this we require
a \textit{candidate generating density}, $q(s,s')$, where $\int
q(s,s') ds' = 1$, which selects the next point in the Markov chain.
The \textit{acceptance probability} of this step is
\begin{equation}
\label{eq: acceptprob} \alpha(s,s') = \mathrm{min} \left( \frac{\pi(s') q(s',s)}{\pi(s) q(s,s')},1 \right ).
\end{equation}
The Metropolis-Hastings algorithm can be summarised as follows
\begin{enumerate}
\item repeat for $i = 1,2,\ldots,n$
\item generate $s'$ from $q(s_i,\cdot)$ and a uniform random variable
  $u$ between 0 and 1.
\item if $u \leq \alpha(s_i,s')$ then set $s_{i+1} = s'$;
\item else set $s_{i+1} = s_i$;
\item return $\{ s_i | i = 1, 2, \ldots, n \}$
\end{enumerate}
The Metropolis-Hastings algorithm generates a Markov chain in the
state space with a transition density $p_{mh}(s'|s) =
q(s,s')\alpha(s,s')$.  When a proposed move is rejected the chain
remains where it is.  This choice of transition density ensures that
the condition of detailed balance is satisfied and hence the Markov
chain converges to the required target density.  The algorithm
generates a Markov chain that finds regions in state space where the
probability distribution is peaked and samples these region with the
correct frequency.

The efficiency of the algorithm is dependent on the choice of
candidate generating function.  Indeed the optimal choice of $q(s,s')$
for a particular target density remains an active area of research
today.  An important special case occurs if the candidate generating
distribution is symmetric, $q(s,s') = q(s',s)$, and so
\begin{equation}
\alpha(s,s') = \mathrm{min} \left( \frac{\pi(s')}{\pi(s)},1 \right).
\end{equation}
This was Metropolis's original formulation which was generalised to
the case of an asymmetric candidate generating function in 1970 by
Hastings~\cite{Hastings1970a}.

\subsection{Estimating sampling error}

In general we wish to determine the weighted average over all possible
states of some observable of the system $$\langle O \rangle_\pi = \int
ds \pi(s) O(s).$$Because the Metropolis algorithm produces a set
samples of the distribution $\pi(\cdot)$, $\{s_i | i = b+1 \ldots b+n
\}$, such averages can be estimated as
\begin{equation}
\label{eq: singlemarkovmean} \langle O \rangle_\pi \simeq \frac{\sum_{i = b+1}^{b+n} O(s_i)}{n}.
\end{equation} 
The statistical uncertainty in such a quantity can be difficult to
estimate because the samples are generally correlated with one
another, as the samples produced by the Metropolis algorithm are not
independent. One way to obtain less correlated samples is to only take
every $g^{th}$ point in the chain after the burn-in for the purpose of
calculating averages, where $g$ is known is the \textit{gap}.
However, this does not usually produce more accurate estimates of
averages than could have been obtained by simply taking every sample
after the burn-in.

Alternatively, one can accept the fact that the samples are
correlated, but attempt to account for the correlations in some
quantitative way.  A scheme for estimating errors in means calculated
from correlated Monte Carlo data may be found in~\cite{Daniell1984a}.

For large dimensional state spaces one may still not be satisfied with
estimating errors from a single Markov chain, even if the correlations
are accounted for. An example is the sampling of probability density
with multiple peaks, which may not individually give correct averages
for state-space dependent quantities.  In this case, the simplest
procedure is to run multiple Markov chains of the same type but with
different starting points, and treat the averages obtained from each
chain as samples of the mean of the quantity of interest.  The overall
mean is calculated by
\begin{equation}
\label{eq: multimarkovmean} \overline{\langle O \rangle} =
\frac{\sum_{i=1}^{\tilde{N}} \langle O \rangle_i}{\tilde{N}},
\end{equation}
where $\langle O \rangle_i$, $i = 1 \ldots \tilde{N}$ are the means
from the independent Markov chains as in Eq.~(\ref{eq:
  singlemarkovmean}), and the subscript $\pi$ has been omitted for
clarity.  By the central limit theorem one would expect samples
obtained in this way to approach a Gaussian distribution about the
true mean.  Therefore an estimate of the error in the mean, for
$\tilde{N}$ independent Markov chains is
\begin{equation}
\label{eq: erromultimarkovmean} \Delta \left[ \overline{\langle  O \rangle} \right ] = \sqrt{\frac{\sum_{i=1}^{\tilde{N}} \left( \langle O \rangle_i -\overline{\langle  O \rangle} \right )^2/\tilde{N}}{\tilde{N} - 1}},
\end{equation} 
where $\Delta[\cdot]$ denotes the error in the mean of the quantity of
interest.  This notation for error in the mean is used throughout this
paper when referring to means obtained from Metropolis data.  

\subsection{Metropolis-Hastings Algorithm for the Stochastic
  Gauge Formalism}

\label{sec: metropolisstochgauge}

The appearance of the weight $\Omega$ as a multiplicative factor in
the stochastic averages in Eq.~(\ref{eq: gaugePquantstochav}) suggests
an interpretation as a probability distribution.  In this section we
show that it is possible to interpret the weight as a
probability distribution over the space of all noise, and hence apply
the Metropolis-Hastings algorithm.

For the purposes of computer simulation, time is necessarily
discretised.  To simulate the time evolution of a system for a period
$T$ we divide the time domain into $N+1$ points so that each step
forward in time is of length $\Delta \tau = T/N$.  For an $M$ mode
system with the standard set of drift gauges (i.e.\ without the extra
gauges and noises discussed in Sec.~\ref{sec: gaugechoice}) we require
$2M$ Gaussian distributed random numbers for each step forward in
time, and thus $2 M N$ random numbers to evolve an entire trajectory.
Hence our fundamental object is a $2MN$- component vector of Gaussian
distributed random numbers $\vec{w} \in \mathbb{R}^{2MN}$, called
the \textit{noise vector}.  The different realisations of $\vec{w}$
give rise to different stochastic trajectories.  The values of all
stochastic phase space variables after $N$ time steps are thus
functions of $\vec{w}$
\begin{equation*}
\vec{\alpha} = \vec{\alpha}[\vec{w}], \quad \Omega = \Omega[\vec{w}].
\end{equation*}
The stochastic average of some some quantity, $\langle O
\rangle_{\rm stoch}$ is
\begin{equation*}
\langle O \rangle_{\rm stoch} = \int d^{2MN} \vec{w} P(\vec{w}) O(\vec{w}) = \lim_{n \rightarrow \infty} \frac{\sum_{i=1}^n O[\vec{w}_i]}{n},
\end{equation*}
where the $\vec{w}_i$ are drawn from a multi-dimensional Gaussian
normal distribution\footnote{Other noise distributions are possible as
  the central limit theorem ensures Gaussian statistics in the limit
  of infinitesimal step size --- however we use Gaussian statistics
  here.}
\begin{equation}
\label{eq: multidimnoise} P(\vec{w}) = \frac{1}{(2 \pi)^{MN}} \exp \left(- \frac{\vec{w}^2}{2} \right),
\end{equation}
and $O = O[\vec{w}]$ is some quantity depending on the noise.

Hence the stochastic sampling of moments in stochastic gauge
simulations, Eq.~(\ref{eq: gaugePquantstochav}), is the
sampling of a multi-dimensional integral
\begin{equation}
\langle \Omega(\vec{w}) O_{mn}(\vec{w}) \rangle_{\rm stoch.} = \int d^{2MN}
\vec{w} P(\vec{w}) \Omega (\vec{w}) O_{mn}(\vec{w}),
\end{equation}
where $O_{mn}(\vec{w}) = \beta[\vec{w}]^m \alpha[\vec{w}]^n + (
\beta[\vec{w}]^n \alpha[\vec{w}]^m )^* $.

The Metropolis-Hastings algorithm can be applied by identifying the
state space $S$ as $\mathbb{R}^{2MN}$ and the state $s$ as the vector
of noises $\vec{w}$.  The probability distribution we wish to sample
is
\begin{equation}
\pi(\vec{w}) = \frac{P(\vec{w}) \Omega [\vec{w}]}{\mathcal{N}},
\end{equation}
where $\mathcal{N} = \int d^{2MN} \vec{w} P(\vec{w}) \Omega
[\vec{w}]$ is a normalisation constant.  
The Metropolis algorithm can be used to generate a set of
samples of $\pi(\vec{w})$, $\left \{ \vec{w}_i | i = 1 \ldots n
\right \}$ and estimate quantum averages as
\begin{equation}
\langle \left( \hat{a}^{\dagger} \right) ^m \hat{a}^n \rangle_{\mathrm{QM}} = \frac{\mathcal{N} \langle O_{mn} \rangle_\pi}{ 2 \mathcal{N}} = \frac{\langle O_{mn} \rangle_\pi}{2} \simeq \frac{\sum_{i=1}^n O_{mn}[\vec{w}_i]/n}{2}.
\end{equation}
None of these steps require explicit knowledge of the normalisation
constant $\mathcal{N}$, although in real-time unitary calculations we
know analytically that it should always be one.  Typically we run
multiple Markov chains as described in Sec.~\ref{sec: metropolis} to
obtain means and standard-deviation error estimates, $\langle O_{mn}
\rangle_\pi = \overline{\langle O_{mn} \rangle} \pm \Delta \left[
  \overline{\langle O_{mn} \rangle} \right ]$.

For comparison, ordinary stochastic sampling uses a Gaussian normal
random number generator to generate samples, $\left \{ \vec{w}_i | i =
  1 \ldots N \right \}$, of $P(\vec{w})$ and calculates quantum
averages as
\begin{equation*}
\langle \left( \hat{a}^\dagger \right)^m \hat{a}^n \rangle_{\mathrm{QM}} \simeq \frac{\sum_{i=1}^n \Omega [\vec{w}_i] O_{mn}[\vec{w}_i]/n}{2}.
\end{equation*}

Finally we note that the Metropolis-Hastings algorithm as outlined
above only optimises the stochastic sampling of moments at the
final \textit{target time}.  At earlier times the probability
distribution $\bar{\pi}(\bar{\vec{w}})$ (where $\bar{\vec{w}}$ is
the noise vector $\vec{w}$ truncated at the point corresponding to
the earlier time) is different in that $\bar{\pi}(\bar{\vec{w'}})/
\bar{\pi}(\bar{\vec{w}}) \neq \pi(\vec{w'}) / \pi(\vec{w})$, and
so we should not expect the Metropolis algorithm targeted to the later
time to correctly sample the moments at earlier times.  This is also
illustrated in Fig.~\ref{fig: weightspread} where the relative weights
vary considerably as time progresses.

\subsection{Designing a Candidate Generating Function}

\label{sec: metropcandfun}

The choice of candidate generating function $q(\vec{w},\vec{w}')$
is a subtle problem and perhaps the most crucial element of applying
the Metropolis-Hastings algorithm to the stochastic gauge formalism.
The Metropolis algorithm generates a Markov chain in noise space,
where at each step the candidate generating function proposes a new
noise $\vec{w}'$ for the stochastic trajectory which is then
accepted or rejected based on the acceptance probability defined by
Eq.~(\ref{eq: acceptprob})
\begin{equation*}
\alpha(\vec{w},\vec{w}') = \mathrm{min} \left(1,\frac{\pi(\vec{w}')
  q(\vec{w}',\vec{w})}{\pi(\vec{w}) q(\vec{w},\vec{w}')} \right) = \mathrm{min} \left(1,\frac{P(\vec{w}') \Omega(\vec{w}') q(\vec{w}',\vec{w})}{P(\vec{w}) \Omega(\vec{w}) q(\vec{w},\vec{w}')} \right).
\end{equation*}
The evaluation of the weight at the target time for the proposed noise
$\Omega(\vec{w}')$ is a non-local procedure; even if only a single
noise is altered in the time domain the entire trajectory from that
time on has to be evolved until the target time in order to evaluate
the new weight.

To separate the issue of sampling high weight trajectories from the
Gaussian nature of the noise it is advisable to choose a candidate
generating function such that
\begin{equation}
\frac{q(\vec{w},\vec{w}')}{q(\vec{w},\vec{w}')} = \frac{P(\vec{w})}{P(\vec{w}')}.
\end{equation}
For this class of generating functions, the probability of a move
being accepted is
\begin{equation}
\alpha(\vec{w},\vec{w}') = \frac{\pi(\vec{w}') q(\vec{w},\vec{w}')}{\pi(\vec{w}) q(\vec{w},\vec{w}')} = \frac{\Omega(\vec{w}') P(\vec{w}') q(\vec{w},\vec{w}')}{\Omega(\vec{w}) P(\vec{w}) q(\vec{w},\vec{w}')} = \frac{\Omega(\vec{w}')}{\Omega(\vec{w})},
\end{equation}
which is only dependent on the weight rather than the Gaussian
distribution, $P(\cdot)$.

\subsection{Time and frequency domain noise functions}

A simple generating function of this type is selecting a
number of entries in the current noise vector and generating new
noises for these.  This procedure can be carried out in either the
time or the frequency domain of the noise.  If the noises are altered
in the time domain the stochastic trajectory is clearly unaltered up
until the point of the first change.  If noises are altered in the
frequency domain then every noise in the time domain is affected to
some extent.  We consider altering noises in the frequency domain to
be a more natural ``small-step'' for the Metropolis algorithm as all
time-domain noises are changed by a small amount on average.  All
results presented in this paper use a generating function operating in
the frequency domain.

In more detail, consider the $R$ independent noise vectors of length
$N$, $\vec{w}_j \in \mathbb{R}^{N}, j = 1 \ldots R$, separately,
instead of a single noise vector of length $RN$.  Insight into
strategies for altering noises may be gained by taking the discrete
Fourier transform of each noise vector
\begin{equation}
\vec{K}_j(n) = \sum_{k=0}^{N-1}e^{2 \pi i k n/N} \vec{w}_j(k).
\end{equation}   
Here a subscript is used to denote each of the $R$ noise vectors, and
should not be confused with the notation $\vec{w}_i$ used previously
to denote the sequence of noise vectors in a Markov chain.  Brackets
$(\cdot)$ denote the components of the noise/spectrum vectors.  The
$n=0$ component of $\vec{K}$ is the mean of the noise, and the fact
that the noise is real places constraints on the components.  Let
$\mathcal{N}(\mu,\sigma)$ denote a Gaussian distribution of mean $\mu$
and standard deviation $\sigma$.  The Fourier transform of a vector
whose components are normally distributed real variables
($\vec{w}(n)$ distributed as $\mathcal{N}(0,1)$) is complex.  For
even $N$ the real and imaginary parts of each component in the range
$2 \leq n \leq N-1$ are distributed as $\mathcal{N}(0, \sqrt{N/2})$
and subject to the constraint $\vec{K}(N-n) = \vec{K}(n)^*$, while the
$n=1$ and $n=N$ components are real and distributed as
$\mathcal{N}(0,\sqrt{N})$.  For odd $N$, the real and imaginary parts
of each component in the range $2 \leq n \leq N$ are distributed as
$\mathcal{N}(0, \sqrt{N/2})$ and subject to the constraint
$\vec{K}(N-n) = \vec{K}(n)^*$, while only the $n=1$ component is real
and distributed as $\mathcal{N}(0,\sqrt{N})$.  These constraints
preserve the total number of independent components $N$.

\begin{figure}[]
\begin{center}
  \centerline{\includegraphics[width=120mm]{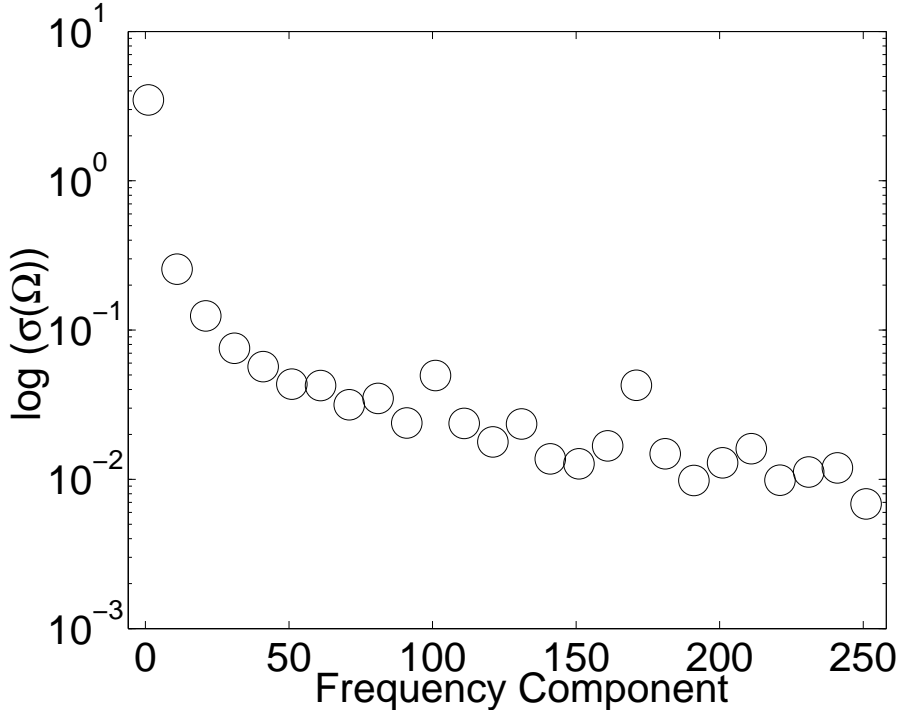}}
\caption{
  Illustration of the effect of varying frequency components of the
  noise vectors on the final value of the weight.  An initial
  stochastic trajectory from a real-gauge simulation of the anharmonic
  oscillator with mean atom number $\bar{n} = 100$ was taken as a
  starting point.  The value of the weight at the end of the
  trajectory, $T= 0.5 /\sqrt{\bar{n}} = 0.05$, was $\Omega = 0.9936$.
  The time step was $\Delta \tau = 10^{-4}$ so each noise
  vector contained 500 components.  In frequency space each noise
  vector contains 250 complex components.  A new random noise was
  chosen for a particular component of the noise vector, the entire
  trajectory re-evolved and the new final weight recorded.  This
  procedure was repeated $10^4$ times for each frequency component and
  the variance in the final weight, $\sigma(\Omega)$, calculated.
  This variance is plotted versus frequency component to give a
  measure of the potential of each frequency component to affect the
  final weight.  Clearly low-frequency noise has more effect than
  high-frequency noise.}
\label{fig: varynoise}
\end{center}
\end{figure}

It is informative to study an individual trajectory and alter noise
elements individually in frequency space to gauge the effect on the
final weight.  One might expect that the low frequency noise should be
more important than the high frequency noise, as high frequency noise
should ``average out'' over a shorter time scale and thus not affect
the dynamics as much.  We find that this is true and Fig.~\ref{fig:
  varynoise} quantifies the potential of frequency components of the
noise to affect the final value of the weight.  To obtain this graph
four random noise vectors were generated and the corresponding
stochastic trajectory evolved using real gauge discussed in
Sec.~\ref{sec: gaugechoice} with mean atom number, $\bar{n}=100$, and
diffusion gauge $A = 2$.  Using this initial trajectory as a starting
point, a particular component of the noise spectrum was examined for
its effect on the final weight by choosing $10^3$ random values for
that component and evolving the corresponding stochastic trajectory,
with all other components of the noise spectrum the same.  In
Fig.~\ref{fig: varynoise} the standard deviation of the final weights
obtained in this manner is plotted on a log scale against the spectrum
component number that was altered.  It is clear from this figure that
the low frequency noises have greater potential to affect the final
weight.  This trend is independent of the initial trajectory, however
we note the actual values of the standard deviations obtained vary
considerably depending on the initial trajectory.

Another question to consider is how many sites in the frequency domain
to alter in proposing a new noise vector, and how to select these.  As
the low-frequency noise seems to affect the final value of the weight
more than the high-frequency noise, one might be tempted to consider
strategies where low-frequency noise is altered more often so as to
more effectively explore the noise space.  In practice we found that
such strategies were no more effective than selecting sites randomly.
Typically we chose to alter of the order $1-10 \%$ of the total
components of the noise.  A good guiding principle is that Metropolis
sampling is thought to be most efficient when approximately $50 \%$ of
proposed moves are accepted during
sampling~\cite{Gilks1996a,Landau2000a}.  We therefore experimented
with different percentages with this principle in mind.

\subsection{Results}

\label{sec: metropolisresults}

We now present the results of a Metropolis sampling of the real-time
dynamics of the anharmonic oscillator using the real gauge discussed
in Sec.~\ref{sec: gaugechoice} and the candidate generating functions
discussed in Sec.~\ref{sec: metropcandfun}. To allow for the
distribution $\pi(\vec{w}) = P(\vec{w}) \Omega(\vec{w})/\mathcal{N}$
to be multiply-peaked we use the more robust technique of estimating
means and errors using multiple Markov chains as discussed in
Sec.~\ref{sec: metropolis}.

\begin{figure}[]
\begin{center}
  \centerline{\includegraphics[width=140mm]{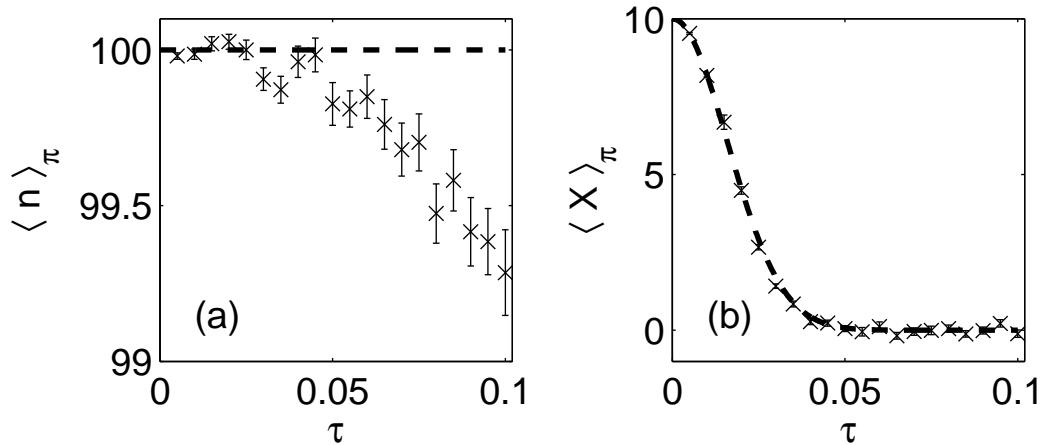}}
 \caption{
  Metropolis sampling of real-gauge simulation of anharmonic
  oscillator dynamics for an initial coherent state with mean atom
  number $\bar{n} = 100$ ($A = 2$, $\lambda = 1/2$).  (a) Mean atom
  number, (b) mean $X-$quadrature.  Metropolis averages and error
  estimates (crosses with error bars) were calculated from the means
  of $10^2$ Markov chains, each taking $10^4$ samples after a burnin
  of $10^4$ proposals.  The total simulation time was $T = 1 /
  \sqrt{\bar{n}} = 0.1$ and the time step used was $\Delta \tau = T/
  10^3 = 10^{-4}$.  Proposals were generated by uniformly selecting
  $10 \%$ of the components of the noise vectors in the frequency
  domain and generating new random noise for those sites.  The
  analytic results are shown as dotted green lines.}
\label{fig: realgaugemetrop}
\end{center}
\end{figure}

Figure~\ref{fig: realgaugemetrop} shows the Metropolis sampling of
$\langle \hat{n} \rangle = \langle \hat{a}^\dagger \hat{a} \rangle $
and $\langle \hat{X} \rangle = \langle (\hat{a} + \hat{a}^\dagger)/2
\rangle$ for a real-gauge simulation of the anharmonic oscillator with
$\bar{n} = 100$.  We targeted the Metropolis sampling to 20 time
points in intervals of $\Delta \tau = 0.005$ from 0 to the final time
of $\tau =T= 1/\sqrt{\bar{n}} = 0.1$.  The average at each time point
is completely independent of the other time points.  Altogether $n =
10^6$ samples were used for the average at each point so statistically
the sampling is comparable to the stochastic sampling in
Fig.~\ref{fig: realgaugestoch}.  However because the Metropolis
sampling has to be run independently for each time point and each
Markov chain has to burn in before sampling begins, there is clearly a
much greater computational effort required to obtain the Metropolis
results.  Nevertheless the Metropolis results are far more reliable
than the stochastic results --- at most time points the average
$\langle \hat{X} \rangle$ is correct to within the estimated error.
In contrast, the stochastic sampling exhibited systematic errors that
were not accounted for by appropriately-large error bars.

We note that even with the Metroplis algorithm the sampled mean atom
number appears to decay slightly at longer times.  Although the
results are still accurate to within 1 \% the estimated error bars do
not account for the difference from the analytic result.  This
behaviour, which is seen for the branching algorithm is well, is due
to the inherent difficulty in sampling the skewed weight distribution
even with Monte Carlo algorithms.  The results are nonetheless vastly
improved compared to stochastic sampling

We experimented with different distributions of the $10^6$ allowed
samples between different numbers of Markov chains.  There is a
trade-off between number of Markov chains and the length of the chain
--- if longer chains are run (e.g.\ $10^2$ chains each with $10^4$
samples) then the means obtained from each chain are more accurate,
and closer to Gaussian about the true mean. However, there are less
chains to average over compared with a run of a larger number of
shorter chains (e.g.\ $10^3$ chains each with $10^3$ samples).  It was
a matter of empirical observation to determine the best balance.

We also experimented with the fraction of noises to be altered in the
frequency domain when proposing a new noise vector.  As a rough guide
we aimed to have $50 \%$ of proposed moves accepted during sampling.
However at short times we found that even if a very large fraction of
the noises were altered (e.g.\ $50 \%$ or more) a large fraction of
proposed changes were accepted ($\sim 90 \%$).  This large acceptance
rate is due to the distribution of weights being relatively narrow at
short times.  At longer times the distribution has spread out
sufficiently such that lower acceptance rates are possible.

Clearly there are many parameters to be optimised in Metropolis
sampling of stochastic gauge equations and we have only scratched the
surface.  We have aimed to present conceptually-simple approaches to
illustrate the principle rather than exhaustively optimise all
parameters.  The fact that enormous improvements were seen over
stochastic sampling --- even with our very simple Metropolis schemes
--- gives us confidence that further improvements are possible.

\section{Branching Algorithm}

\label{sec: branching}

The second Monte Carlo technique that we investigate for real-time
stochastic-gauge simulations is a branching algorithm similar to that
used in Green's function Monte Carlo, see e.g.\ \cite{Trivedi1990a}.
Corney and Drummond~\cite{Corney2004a} have previously used this
algorithm for stochastic simulations in imaginary time using a
Gaussian basis (a generalisation of the stochastic-gauge basis).

The branching algorithm is simpler to describe and more
straightforward to apply to stochastic gauge simulations.  Another
advantage is that there are fewer free parameters than the Metropolis
algorithm.  The branching algorithm works by concurrently evolving a
``population'' of stochastic trajectories in time, and periodically
cloning those that acquire a large weight and killing those that
acquire a small weight.  

\subsection{Algorithm}

We define $T$ to be the total simulation time, $\Delta
\tau_{\rm b}$ be the time interval between branching events and
$\Delta \tau$ the fundamental time-step for integrating the SDEs.  In
practice it is desirable for the number of branching events $B =
T/\Delta \tau_{\rm b}$ and the number of time steps in a branching
period $\Delta \tau_{\rm b} / \Delta \tau$ to be integers.  Formally
the branching algorithm can be stated as
\begin{enumerate}
\item Begin with an initial population of $N_{\mathrm{pop}}$
  stochastic trajectories.
\item Evolve all stochastic trajectories forward in time for a period
  $\Delta \tau_{\rm b}$
\item for i = 1, 2, \ldots, $N_{\rm pop}$
\begin{itemize}
\item Generate a uniform random variable $u$ between 0 and 1
\item Make $m_i = {\rm int}[\Omega_i/\overline\Omega + u]$ clones of
  trajectory $i$.
\item Set $\Omega_i = 1$
\end{itemize}
\item Set $N_{\rm pop} = \sum_{i = 1}^{N_{\rm pop}} m_i$.
\end{enumerate}
Here we set $\overline\Omega = \langle \Omega \rangle$ to ensure that
the number of trajectories in the population $N_{\rm pop}$ remains
constant on average.  Because $\Omega$ does not couple into the SDEs
for the mode variables, further evolution is not affected by resetting
of the weights at each branching time.  The statistical equivalence
between the weight and the multiplicity of paths means that the
physical moments are unchanged on average by the branching procedure.
To see this note that the average of the mean of an observable $O$
after the branching event is
\begin{eqnarray}
\langle O \rangle &=& \left. \sum_{i = 1}^{N_{\rm pop}} \int_0^1 m_i(u)du\,O_i\right\slash \sum_{i = 1}^{N_{\rm pop}}\int_0^1 m_i(u)du \nonumber\\
&=& \left. \sum_{i = 1}^{N_{\rm pop}} \left(\Omega_i/\overline\Omega\right) O_i\right\slash\sum_{i = 1}^{N_{\rm pop}} \Omega_i /\overline\Omega\nonumber\\
&=& \frac{1}{N_{\rm pop}} \sum_{i = 1}^{N_{\rm pop}} \Omega_i O_i, 
\end{eqnarray}
which is identical to the mean before the branching event.

\subsection{Results}

In this section we present the results of a branching-algorithm
sampling of real-time dynamics of the anharmonic oscillator using the
real gauge discussed in Sec.~\ref{sec: gaugechoice}.  Similarly to the
Metropolis sampling, we calculate averages and error estimates from
multiple independent populations.

\begin{figure}[]
\begin{center}
 \centerline{\includegraphics[width=140mm]{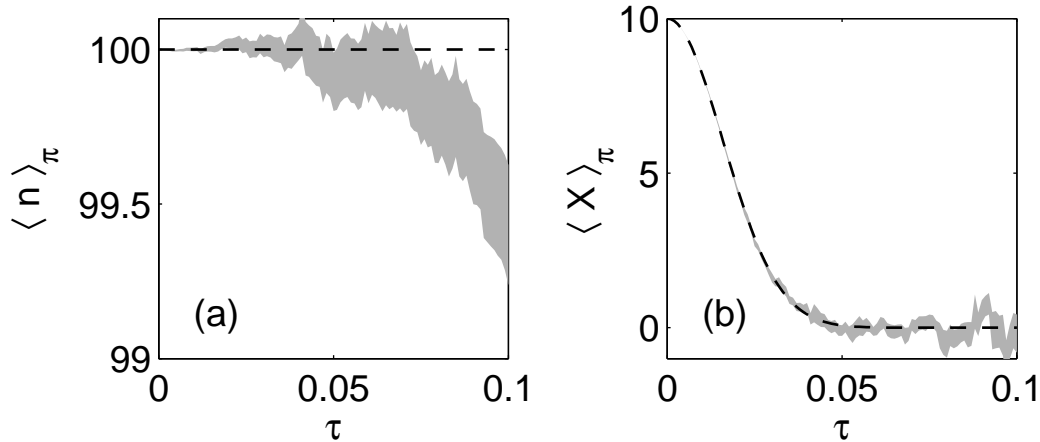}}
\caption{
  Branching-algorithm sampling of real-gauge simulation of anharmonic
  oscillator dynamics for an initial coherent state with mean atom
  number $\bar{n} = 100$ ($A = 2$, $\lambda =1/2$).  (a) Mean atom
  number, (b) mean $X-$quadrature.  Branching averages and error
  estimates, indicated by the shaded region, were calculated from the
  means of $10^2$ populations, each containing $10^4$ trajectories on
  average.  The total simulation time was $\tau = 1/\sqrt{\bar{n}} =
  0.1$ and the time step used was $\Delta \tau = \tau / 10^3 =
  10^{-4}$.  The branching time was $\Delta \tau_{\rm b} = 10^{-3}$.
  The analytic results are shown as dotted lines.}
\label{fig: realgaugebranch}
\end{center}
\end{figure}

Fig.~\ref{fig: realgaugebranch} shows a branching-algorithm sampling
of $\langle \hat{n} \rangle = \langle \hat{a}^\dagger \hat{a} \rangle
$ and $\langle \hat{X} \rangle = \langle (\hat{a} + \hat{a}^\dagger)/2
\rangle$ for a real-gauge simulation of the anharmonic oscillator with
$\bar{n} = 100$.  Again we see an enormous improvement in the sampling
compared with stochastic sampling, Fig.~\ref{fig: realgaugestoch}.  A
clear advantage of the branching algorithm over Metropolis is that it
generates physical moments at every time step as opposed to being
targeted to a single time.

At each time there are, on average, $10^6$ stochastic trajectories
contributing to the stochastic averages. As with the Metropolis, it
was again a matter of experimenting with different ratios of the
number of trajectories to the number of populations to determine the
best balance.  We present results for the same division of
trajectories amongst populations ($10^2$ populations of $10^4$
trajectories) as samples amongst Markov chains for the Metropolis
algorithm so that the results are as comparable as possible.

The only free parameter in the algorithm itself is the time between
branching events, $\Delta \tau_{\rm b}$.  There is a trade-off between
making this interval large enough that the weights spread out
sufficiently for the branching to be meaningful and small enough to
improve the sampling continuously in time.  The branching interval
used for this simulation, $\Delta \tau_{\rm b} = 10^{-3}$, is small on
the scale of the dynamics of the system but large enough to allow the
weights to diverge significantly between branching events.

\section{Conclusions and Outlook}

\label{sec: conclusion}

In this paper we have demonstrated the use of two Monte Carlo
techniques, the Metropolis algorithm and a branching algorithm, for
real-time calculations of quantum dynamics with the stochastic-gauge
method.  This work should be considered a proof of principle rather
than a fully optimised `recipe'.  It is part of a larger program of
optimising bases, gauges and algorithms for stochastic simulations of
quantum dynamics in real and imaginary time.  A timely application of
these methods is in theoretical calculations for ultra-cold atomic
gases~\cite{Drummond2004a,Corney2004a}. QMC methods have been used to
calculate some static properties of ultra-cold gases, e.g. see
\cite{Carlson2003a,Astrakharchik2004a}.  These systems are
quantum-many body by nature and hence few exact theoretical results
exist.  They are an ideal testing ground for theory due to their
purity and well-understood controllable interactions.

In this work we have restricted ourselves to real gauges so that the
weights remain real and can be interpreted as probabilities, and
have considered the single-mode anharmonic oscillator as an example of
our methods.  In order to control the divergence of stochastic
trajectories using real gauges we have explored a previously untested
gauge freedom resulting from the choice of a non-square noise matrix.
The resulting distribution of weights becomes highly skewed on a time
scale proportional to the inverse of the mean atom number.  The weight
parameter for most stochastic trajectories tends towards zero, whereas
very few tend towards a large weight.  Such distributions are likely
to be ubiquitous for unitary real-time stochastic-gauge simulations
with real gauges, as the weight distribution necessary broadens with
time while the mean weight must remain unity.

While such skewed distributions are difficult to sample with the usual
stochastic methods, they are ideally suited to Monte Carlo importance
sampling techniques that preferentially sample high-weight
trajectories.  Indeed we found enormous improvements over stochastic
sampling using both the Metropolis and branching algorithms.  The
branching algorithm is the more straightforward to apply because it
has only one free parameter (the branching interval) and produces
results at every time step.  We suggest it as the best starting point
for future Monte Carlo simulations.  By contrast the Metropolis
algorithm has to be targeted to a particular time and so seems less
useful.  However, there is a lot more freedom in the Metropolis
algorithm and a vast literature exists on optimising sampling for
particular distribution.  Thus it seems quite possible that it will be
better suited to some problems, especially when further improvements
over the branching algorithm are desirable.

Traditionally Monte Carlo techniques have been highly successful in
imaginary-time calculations for thermal equilibrium.  This paper has
extended the use of these techniques to real-time quantum-dynamical
calculations and thus opens a new domain of application for these
algorithms. In future, these techniques need to be extended to
many-mode, many-particle problems where exact solutions are not known.

\section*{Acknowledgments}
  
We would like to thank Piotr Deuar for helpful discussions, and the
Australian Research Council, the Queensland state government and the
University of Queensland for financial support.

\end{document}